\documentclass[aps,prc,preprint,amsmath,amssymb,showpacs,preprintnumbers,superscriptaddress]{revtex4-1}
\usepackage{CJK}
\usepackage{graphicx}
\usepackage{dcolumn}
\usepackage{bm}
\usepackage{color}
\usepackage{hyperref}

\allowdisplaybreaks

\begin{document}
\begin{CJK*}{GB}{}
\title{Time-dependent density functional theory study of induced-fission dynamics of $^{226}$Th}

\CJKfamily{gbsn}
\author{B. Li \CJKfamily{gbsn} (À)}
\affiliation{State Key Laboratory of Nuclear Physics and Technology, School of Physics, Peking University, Beijing 100871, China}
\author{D. Vretenar}
\email{vretenar@phy.hr}
\affiliation{Physics Department, Faculty of Science, University of Zagreb, 10000 Zagreb, Croatia}
\affiliation{State Key Laboratory of Nuclear Physics and Technology, School of Physics, Peking University, Beijing 100871, China}
\author{T. Nik\v si\' c}
\affiliation{Physics Department, Faculty of Science, University of Zagreb, 10000 Zagreb, Croatia}
\affiliation{State Key Laboratory of Nuclear Physics and Technology, School of Physics, Peking University, Beijing 100871, China}
\author{P. W. Zhao \CJKfamily{gbsn} (ÕÔÅôΡ)}
\email{pwzhao@pku.edu.cn}
\affiliation{State Key Laboratory of Nuclear Physics and Technology, School of Physics, Peking University, Beijing 100871, China}
\author{J. Meng \CJKfamily{gbsn} (ÃϽÜ)}
\email{mengj@pku.edu.cn}
\affiliation{State Key Laboratory of Nuclear Physics and Technology, School of Physics, Peking University, Beijing 100871, China}

\begin{abstract}
A microscopic finite-temperature model based on time-dependent nuclear density functional theory (TDDFT), is employed to study the induced-fission process of $^{226}$Th. The saddle-to-scission dynamics of this process is explored, starting from various points on the deformation surface of Helmholtz free energy at a temperature that corresponds to the experimental excitation energy, and following self-consistent isentropic fission trajectories as they evolve toward scission. Dissipation effects and the formation of excited fragments are investigated and, in particular, the difference in the evolution of the local temperature along asymmetric and symmetric fission trajectories. The relative entropies and entanglement between fission fragments emerging at scission are analyzed.
\end{abstract}

\date{\today}

\maketitle

\end{CJK*}
\section{Introduction}
Nuclear fission is a decay process by which a heavy nucleus splits into two or more lighter fragments. In addition to being intrinsically interesting for studies of nuclear dynamics, fission plays an important role in the synthesis of superheavy elements~\cite{Hofmann2000RMP}, and the rapid neutron-capture process of nucleosynthesis in stellar environments~\cite{Korobkin2012MNRAS,Just2015MNRAS}. Even though a large amount of data have been accumulated in experiments over many decades, important
fission observables, such as half-lives, charge and mass yields, total kinetic energy, total excitation energy, and the decay pattern of fission fragments, for many unstable nuclei that have not been produced in laboratory, necessitate accurate theoretical predictions. A number of models, based on both phenomenological and microscopic approaches, have been developed and applied to describe fission dynamics. Phenomenological models often include untested assumptions and, of course, their parameters are adjusted to fission data in one way or another.
In recent years, a significant effort has been made to develop microscopic theories~\cite{Schunck16PPNP,Schunck22PPNP}.
Based on two principal microscopic approaches, that is, the time-dependent density functional theory (TDDFT)~\cite{Ren_22PRL,simenel12,simenel18,nakatsukasa16,stevenson19,bulgac16,magierski17,scamps18,bulgac19,bulgac20}, and the time-dependent generator coordinate method (TDGCM) \cite{krappe12,younes19,Regnier2016_PRC93-054611,Tao2017PRC,Verriere2020_FP8-233},
extensive studies of various aspects of the fission process have been reported. Although it represents a fully quantum mechanical and microscopic approach, the TDGCM describes fission as an adiabatic process and, therefore, does not include any dissipation mechanism and cannot be applied to study the excitation of fission fragments.
Several attempts to include a dissipation mechanism in TDGCM have been made~\cite{bernard11,Dietrich2010NPA,Zhao2022PRC,Zhao2022PRC_TKE}, but the resulting models are rather complex and difficult to use in large-scale calculations of fission properties. The TDDFT models fission events by propagating individual nucleons independently in self-consistent mean-field potentials, and automatically includes the one-body dissipation mechanism. In the standard formulation, TDDFT cannot be used in the classically forbidden region of the collective space, where the potential energy (barrier) is higher than the total energy of the initial state, and does not consider quantum fluctuations in collective degrees of freedom. The description of mean-field dynamics can be improved by including a one-body fluctuation mechanism such as, for instance, in stochastic TDDFT \cite{Ayik2008PLB}, or one can add additional terms to the standard TDDFT equations to simulate both dissipation and fluctuations of nuclear collective motion, while the evolution remains unitary \cite{Bulgac2019PRC}.
An important advantage with respect to the GCM approach, is the capability of TDDFT-based models to describe dissipative effects and excitation of fission fragments.

The dynamics of induced fission is particularly interesting, because observables exhibit a complex dependence on the excitation energy of the fissioning system. For a quantitative description of observables in an induced fission process, one has to explicitly take into account the excitation energy of the compound nucleus in a microscopic framework, or parametrize it in terms of nuclear temperature \cite{Schunck2015PRC}.
Both in the TDGCM framework~\cite{Zhao2019_PRC99-014618,Zhao2019_PRC99-054613,Schunck2020_LLNL-PROC-767225,McDonnell2014_PRC90-021302R,McDonnell2013_PRC87-054327,Martin2009_IJMPE18-861}
and TDDFT~\cite{zhu16,qiang21,PhysRevC.104.054604}, several models have been developed and applied to fission dynamics at finite temperatures. However, these models do not include the local change in nuclear temperature and, thus, cannot describe the evolution of temperature as the fissioning system evolves toward scission. Recently, a TDDFT-based microscopic finite-temperature model has been formulated, that allows to follow the evolution of local temperature along fission trajectories. For few illustrative cases of induced fission characterized by asymmetric charge and mass yield distributions, the partition of the total energy into various kinetic and potential energy contributions at scission has been analyzed \cite{Li2023PRC_FTTDDFT}.

In this work, we employ the newly developed TDDFT finite-temperature framework in a study of induced fission dynamics of $^{226}$Th, a system that exhibits both symmetric and asymmetric peaks (three peaks structure) of the yields distribution  \cite{Schmidt2000NPA,Chatillon2019PRC}. In a previous work \cite{Zhao2019_PRC99-014618}, the charge yields for induced fission of $^{226}$Th were analyzed using the finite-temperature TDGCM in the Gaussian overlap approximation (GOA). It was shown that the precise experimental position of the asymmetric peaks
and the symmetric-fission yield can only be accurately reproduced when the potential and inertia tensor of the collective Hamiltonian are determined at finite temperature that corresponds to the experimental excitation energy. However, the TDGCM model employed in Ref.~\cite{Zhao2019_PRC99-014618} does not include energy dissipation and heating of fission fragments. Here, we focus on dissipation effects and analyze the differences in the evolution of temperature along asymmetric and symmetric fission trajectories. Starting from a temperature that corresponds to the experimental excitation energy of the compound system, the TDDFT model propagates nucleons along isentropic trajectories toward scission. Close to scission, the hexadecapole deformation $\beta_{40}$ provides a measure of the thickness of the neck between the two emerging fragments, and here its influence on fission trajectories is investigated. Following our recent study \cite{Li2024arxiv} on entanglement in multinucleon transfer reactions, in this work we extend the analysis to entropies and entanglement between fragments in symmetric and asymmetric fission. The theoretical framework is outlined in Sec.~\ref{sec_theo}.
The dissipative saddle-to-scission dynamics and properties of the emerging fragments are explored in Sec.~\ref{sec_results}.
Section~\ref{sec_summ} summarizes the principal results of this study.

\section{Theoretical framework}\label{sec_theo}
\subsection{Finite temperature time-dependent density functional theory}
If a compound nucleus is in a state of thermal equilibrium at temperature $T$,
the system can be described by the finite temperature (FT) Hartree-Fock-Bogoliubov (HFB) theory \cite{goodman1981finite}.
In the grand-canonical ensemble, the expectation value of any operator $\hat{O}$ is given by an ensemble average
\begin{equation}
\langle \hat{O} \rangle = \textrm{Tr} ~[ \hat{D}\hat{O} ],
\end{equation}
where $\hat{D}$ is the density operator:
\begin{equation}
\hat{D} = {1 \over \mathcal{Z} } ~ e^{ -\beta \left( \hat{H}-\lambda \hat{N} \right) }\; .
\end{equation}
$\mathcal{Z}$ is the grand partition function, $\beta=1/k_{B}T$ with the Boltzmann constant $k_{B}$, $\hat{H}$ is the Hamiltonian of the system,
$\lambda$ denotes the chemical potential, and $\hat{N}$ is the particle number operator.

In the present study of induced fission dynamics of $^{226}$Th, the internal excitation energy $E^*_{FS}$ of the fissioning system, defined as the difference between the total binding energies of the equilibrium self-consistent mean-field minima at temperature $T$ and at $T = 0$,
corresponds to temperatures that are above the pairing phase transition.
The temperature at which pairing correlations vanish depends on the specifics of the nucleus but, for the case of induced fission of $^{226}$Th, considered in the present work,
the pairing energy is negligible at temperatures $T \geq 0.8$ MeV~\cite{Zhao2019_PRC99-014618}. Then, the corresponding single-particle equation reads
\begin{equation}\label{Eq_Dirac}
    \hat{h}\psi_k = \varepsilon_k\psi_k,
\end{equation}
where, in the relativistic case, the Dirac Hamiltonian $\hat{h}$ is defined by
\begin{equation}
   \hat{h} = \bm{\alpha}\cdot(\hat{\bm{p}}-\bm{V})+V^0+\beta(m_N+S),
   \label{Ham_D}
\end{equation}
with the scalar $S(r)$ and vector fields $V^\mu(r)$, respectively
\begin{subequations}
  \begin{align}
    S=\,&\alpha_S\rho_S+\beta_S\rho_S^2+\gamma_S\rho_S^3+\delta_S\Delta\rho_S,\\
    V^\mu=\,&\alpha_Vj^\mu+\gamma_V(j^\mu j_\mu)j^\mu+\delta_V\Delta j^\mu+\tau_3\alpha_{TV}j_{TV}^\mu+\tau_3\delta_{TV}\Delta j_{TV}^\mu+e\frac{1-\tau_3}{2}A^\mu.
  \end{align}
\end{subequations}
In the absence of pairing correlations at finite temperature $T$, the local densities and currents $\rho_S$, $j^\mu$, and $j_{TV}^\mu$ can be written in the following form:
\begin{subequations}\label{Eq_density_current}
  \begin{align}
    &\rho_S=\sum_{k=1}f_k\bar{\psi}_k\psi_k,\\
    &j^\mu=\sum_{k=1}f_k\bar{\psi}_k\gamma^\mu\psi_k,\\
    &j_{TV}^\mu=\sum_{k=1}f_k\bar{\psi}_k\gamma_\mu\tau_3\psi_k,
  \end{align}
\end{subequations}
where $f_k$ is the thermal occupation probability,
defined as a function of single-particle energy $\varepsilon_k$ in Eq.~\eqref{Eq_Dirac}, the temperature $T$, and chemical potential $\lambda$:
\begin{equation}
  f_k = \frac{1}{1+e^{(\varepsilon_k-\lambda)/k_BT}}.
\label{thermal_f}
  \end{equation}
The chemical potential $\lambda$ is determined numerically in such a way that the particle number condition $\sum_k f_k = N$ is fulfilled.
The entropy of this nucleus can be obtained from the occupation probability,
\begin{equation}\label{Eq_entropy}
S=-\sum_k [f_k \ln f_k +(1-f_k) \ln (1-f_k)].
\end{equation}

In the dynamic case~\cite{Li2023PRC_FTTDDFT}, the evolution of single-nucleon spinors $\psi_k$ is governed by the time-dependent Kohn-Sham equation~\cite{Rung1984TDDFT,Leeuwen1999TDDFT},
\begin{equation}\label{Eq_td_Dirac_eq}
  i\frac{\partial}{\partial t}\psi_k(t)=\hat{h}(t)\psi_k(t).
\end{equation}
The dependence on time of the Dirac Hamiltonian $\hat{h}(t)$ is determined self-consistently by the time-dependent densities and currents~\cite{Rung1984TDDFT}.
The functional dependence of local densities and currents on temperature is the same as in the static case, with the time-dependent thermal occupation $f_k$,
\begin{equation}\label{eq_fkt}
  f_k(t) = \frac{1}{1+e^{[\varepsilon_k(t)-\lambda(t)]/k_BT(t)}}.
\end{equation}
The single-particle energy $\varepsilon_k(t)$ is defined: $\varepsilon_k(t)=\langle\psi_k(\bm{r},t)|\hat{h}(\bm{r},t)|\psi_k(\bm{r},t)\rangle$ and, in this case, both $T(t)$ and $\lambda(t)$ are time-dependent.
The functional dependence of the time-dependent entropy on occupation probabilities is the same as in the static case,
\begin{equation}\label{Eq_entropy}
S(t)=-\sum_k [f_k(t) \ln f_k(t) +(1-f_k(t)) \ln (1-f_k(t))].
\end{equation}
Starting from the initial stationary values, the Lagrange multipliers $\lambda(t)$ and $T(t)$, considered as non-equilibrium generalization of the chemical potential and temperature,
are adjusted at each step in time in such a way that the particle number and total energy are conserved along a self-consistent TDDFT trajectory. As shown in Ref.~\cite{Li2023PRC_FTTDDFT}, this results in an isentropic fission trajectory, that is, the local entropy (\ref{Eq_entropy}) is constant along a trajectory leading to scission.

\subsection{Numerical details}\label{sec:paths}
In calculations with the time-dependent covariant DFT~\cite{Ren_22PRL,ren20LCS,ren20O16,Ren_22PRC,Li2023PRC_gdTDGCM,Li2024FOP},
the mesh spacing of the lattice is 1.0 fm for all directions, and the box size is $L_x\times L_y\times L_z=20\times20\times60~{\rm fm}^3$.
The time-dependent Dirac equation \eqref{Eq_td_Dirac_eq} is solved with the predictor-corrector method.
The step for the time evolution is $6.67\times10^{-4}$~zs.
The point-coupling relativistic energy density functional PC-PK1~\cite{zhao10} is adopted in the particle-hole channel.
The ground state energy of $^{226}$Th at $T=0$ is obtained by the self-consistent relativistic DFT calculations in a three-dimensional (3D) lattice space, with pairing correlations treated in the BCS approximation. A monopole pairing interaction is employed, and the pairing strength parameters: $-0.165$ MeV for neutrons, and $-0.250$ MeV for protons, are determined by the empirical pairing gaps of $^{226}$Th,
using the three-point odd-even mass formula~\cite{Bender2000pairing}.
The initial states for the time evolution at finite temperature, are obtained by self-consistent deformation-constrained relativistic DFT calculations in a 3D lattice space~\cite{Ren2017PRC, ren19LCS, ren20_NPA,Li2020PRC,Xu2024PRC}, with the box size: $L_x\times L_y\times L_z=20\times20\times50~{\rm fm}^3$.
\section{RESULTS AND DISCUSSIONS}\label{sec_results}
\subsection{Asymmetric and symmetric fission trajectories in $^{226}$Th}

\begin{figure}[!htbp]
\centering
\includegraphics[width=0.8\textwidth]{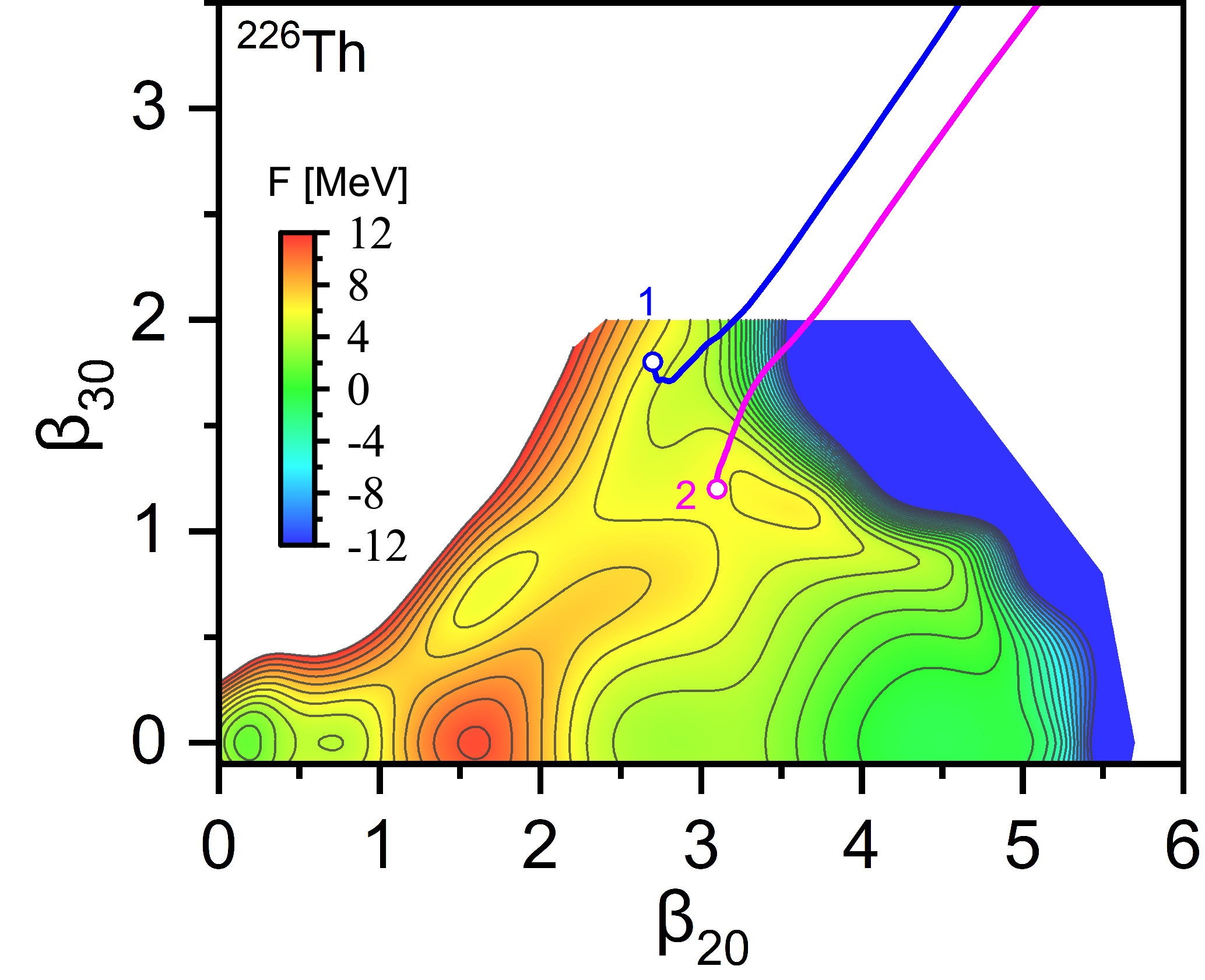}
\caption{Self-consistent deformation surface of Helmholtz free energy $F=E(T)-TS$ of $^{226}$Th in the plane of quadrupole-octupole axially-symmetric deformation parameters,
calculated with the relativistic density functional PC-PK1 at the temperature $T=0.8$ MeV. Contours join points on the surface with the same energy (in MeV), and the contour interval is 1 MeV.
The curves denote the finite temperature TDDFT fission trajectories for two initial points on the energy surface.}
 \label{fig:ES_beta20_beta30}
\end{figure}

The ground-state energy of $^{226}$Th at $T=0$, obtained by the self-consistent relativistic DFT with the energy density functional PC-PK1 and a monopole pairing interaction, is $-1730.47$ MeV, to be compared with the experimental value: $-1730.51$ MeV.
In the experiment of photo-induced fission of $^{226}$Th, the photon energies were in the interval $8-14$ MeV, with a peak value of $E_{\gamma} = 11$~MeV~\cite{Schmidt2000NPA},
and the more recent study \cite{Chatillon2019PRC} measured the elemental yields of $^{226}$Th, with the mean excitation energy $\langle E^{*}\rangle= 13.5$ MeV.
In this work, we analyze the dynamics of induced fission of $^{226}$Th at initial temperature $T=0.8$ MeV and internal excitation energy $E^*_{FS} = 10.47$ MeV,
which corresponds to the peak photon energy in the experiment of Ref.~\cite{Schmidt2000NPA}. Figure~\ref{fig:ES_beta20_beta30} displays the self-consistent deformation energy surface (Helmholtz free energy $F=E(T)-TS$) at $T=0.8$ MeV, as a function of two collective coordinates: the axial quadrupole $\beta_{20}$ and octupole $\beta_{30}$ deformation parameters.
The equilibrium minimum is located at $(\beta_{20}, \beta_{30})\approx(0.18,0.13)$, the isomeric minimum is at $(\beta_{20}, \beta_{30})\approx (0.70, 0)$,
and the third local minimum is found at $(\beta_{20}, \beta_{30})\approx (1.71, 0.73)$ in the asymmetric valley. Along the symmetric fission trajectory, i.e., for $\beta_{30}=0$, we find four local minima.

\begin{figure}[!htbp]
\centering
\includegraphics[width=0.9\textwidth]{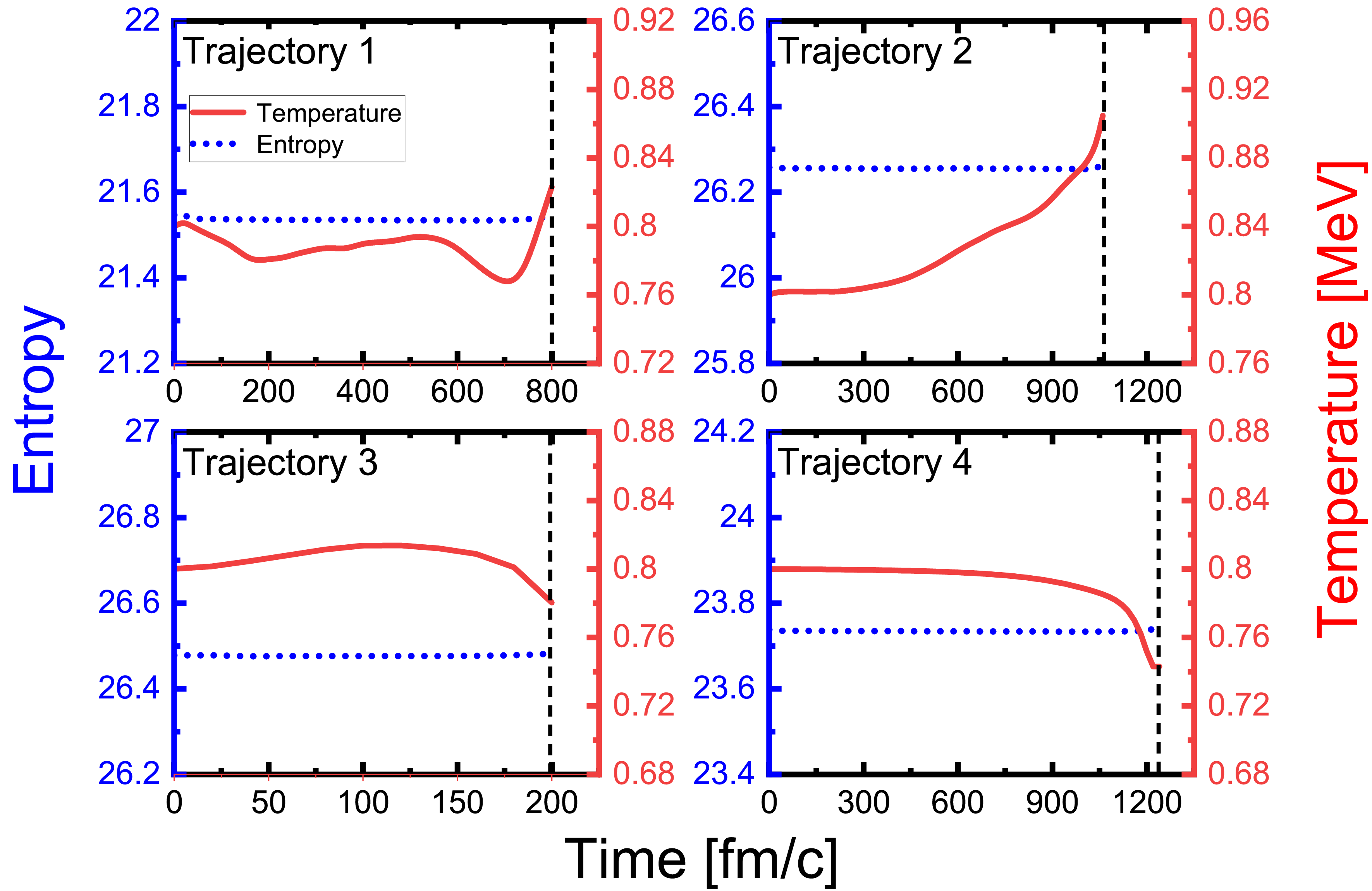}
\caption{Local temperature and entropy as functions of time, for trajectories 1 and 2 shown in Fig.~\ref{fig:ES_beta20_beta30}, and trajectories 3 and 4 in Fig.~\ref{fig:ES_beta20_beta40_sym}. The vertical dashed lines denote the time at scission, starting from the initial point of each trajectory.}
 \label{fig:entropy_temperature}
\end{figure}

The open dots denote two points on the deformation energy surface, beyond the outer fission barrier (saddle point), that will be used as initial points for the illustrative calculation of asymmetric fission trajectories. It is well known that, since TDDFT describes a classical evolution of independent nucleons in mean-field potentials, it cannot be applied in the classically forbidden region of the collective space and, therefore, the initial locations for TDDFT calculations have to be selected beyond the outer fission barrier. The trajectories 1 and 2 in Fig.~\ref{fig:ES_beta20_beta30} start at the deformations $(\beta_{20},\beta_{30})=(2.70,1.80)$ and $(\beta_{20},\beta_{30})=(3.10,1.20)$ respectively. The time-evolution of the corresponding local temperatures and entropies are shown in Fig.~\ref{fig:entropy_temperature}, from the initial points to scission. Even though TDDFT propagates nucleon wave functions also beyond scission, the current implementation of the model cannot take into account different temperatures of the fission fragments. The initial temperature corresponds to the excitation energy of the full fissioning system. However, after scission, the two fragments evolve independently and eventually relax to their equilibrium shapes, with their deformation energies converted into intrinsic excitations. This means that the temperature of each fragment evolves independently, but the model cannot consider  different temperatures in the two regions of space in which the fragments travel.
As explained in the previous section, since the local temperature is adjusted at each step in time in such a way that the total energy is conserved, the resulting trajectory is isentropic, that is, the entropy remains constant in the closed system on the trajectory to scission.

\begin{table}[!htbp]
  \caption{Initial temperature $T^{init}$, total energy at the initial point $E_{init}$, and initial excitation energy $E^{*}_{FS}$ of the fissioning system, various components of the total energy at scission, and the final temperature at scission $T_{sci}$, for the fission trajectories 1 - 6 of $^{226}$Th, shown in Figs.~\ref{fig:ES_beta20_beta30} - \ref{fig:beta40_effect}, respectively. All values are given in MeV.}
  \bigskip
  \centering
  \renewcommand\arraystretch{1.5}
   \begin{tabular}{ccccccc}
    \hline\hline
      nucleus &\multicolumn{6}{c}{$^{226}$Th}\\
    \hline
      trajectory & 1 & 2 & 3 & 4 & 5 & 6 \\

     \hline
     $T^{init}$               & 0.80     & 0.80     &  0.80    &  0.80    &   0.80   &  0.80        \\

     $E_{init}$               & $-1707.15$ & $-1703.12$ & $-1708.53$ & $-1711.48$ & $-1710.64$ & $-1685.31$     \\

     $E^{*}_{FS}$             & 10.47    & 10.47    & 10.47    & 10.47    & 10.47    & 10.47        \\

     $E^{k,pre}$              & 12.19    & 12.56  	&  6.23    & 5.57     & 5.87     & 16.38        \\

     $E^{k,C}$                & 143.60   & 144.18   &  147.18  &  144.79  &  143.67  &  158.61      \\

     $E^{1,T=0}_{g.s.}(M_1)$  & $-1155.21$ & $-1131.11$ & $-946.95$  & $-946.95$  & $-946.95$  & $-946.95$      \\

     $E^{*,def}_1$            & 1.82	 & 11.12    & 8.88     & 9.01	  & 9.01	 &  6.61        \\

     $E^{2,T=0}_{g.s.}(M_2)$  & $-735.74$  & $-767.06$  & $-946.95$  & $-946.95$  & $-946.95$  & $-946.95$      \\

     $E^{*,def}_2$            & 6.16	 & 5.73	    & 8.88     & 9.01	  & 9.01	 &  6.61       \\

     $E^{*,int}$              & 20.03    & 21.47    & 14.20    &  14.04   & 15.70    &  20.38      \\

     $E^{*,dis}$              & 9.56     & 11.00    & 3.73     &  3.57    & 5.23     &  9.91       \\

     $T_{sci}$                & 0.82     & 0.90     & 0.78     &  0.74    & 0.76     &  0.78        \\

     \hline
  \hline\hline
  \end{tabular}
    \label{tab:dissipation}
\end{table}

At first glance, the two trajectories in Fig.~\ref{fig:ES_beta20_beta30} are similar, that is, it appears they simply follow the trajectory of steepest descent toward asymmetric scission. However, they are characterized by distinct values of the entropy, and the evolution of the local temperature is also rather different, as shown in the two top panels of Fig.~\ref{fig:entropy_temperature}. Trajectory 2 has a larger entropy, and it also exhibits a uniform increase of temperature toward scission, at which point its value is enhanced by $\approx 10\%$. For trajectory 1, the temperature initially oscillates because of the exchange between the deformation energy and the intrinsic excitation energy. Close to scission, dissipation becomes more pronounced and this leads to a steep temperature increase to a value that is essentially the same as at the initial point.

The effects of energy dissipation and heating along the trajectories 1 and 2, are summarized in the first two columns of Table \ref{tab:dissipation}. $T^{init}$ is the initial temperature, $E_{init}$ is the total energy at the initial point, and $E^*_{FS}$ the initial  intrinsic excitation energy of the fissioning system. $E^{k,pre}$ is the collective flow energy at scission and $E^{k,C}$ is the Coulomb energy between two fragments at scission.
$E^{1,T=0}_{g.s.}$ and  $E^{2,T=0}_{g.s.}$ are the ground-state binding energies of the fragment 1 and fragment 2, respectively, at $T=0$. $E^{*,def}_1$ and $E^{*,def}_2$ are the corresponding deformation energies at the scission temperature $T_{sci}$. $E^{*,int}$ is the intrinsic excitation energy of the two fragments at scission, and
$E^{*,dis}$ is the energy dissipated along the fission trajectory.
These quantities are determined by the energy balance relation:
\begin{equation}
E_{init} =E^{1,T=0}_{g.s.}+ E^{2,T=0}_{g.s.} +TKE +TXE
\label{eq:en_balance}
\end{equation}
The total kinetic energy TKE consists of the Coulomb energy $E^{k,C}$ between the fragments at scission, and the pre-scission kinetic energy $E^{k,pre}$ which results from a partial conversion of the saddle-to-scission collective potential energy difference (the other part is converted into the deformation energy of the fragments and dissipation energy). $E^{k,pre}$ is the collective flow energy at scission
\begin{equation}
E^{k,pre} = \frac{m}{2} \int   \rho(\vec{r}, t_{sci}) {\vec{v}}~^2 (\vec{r},t_{sci}) d\vec{r}\;,
\label{eq:en_flow}
\end{equation}
where the density and velocity field are evaluated at the time of scission.
The total excitation energy TXE is divided into the deformation energy of the fragments at scission and the intrinsic excitation energy,
\begin{equation}
TXE = \sum_{i=1}^2 E_i^{*,def} + E^{*,int}\; .
\label{eq:en_INT}
\end{equation}
The former can be easily computed by taking, for each fragment at $T=T_{sci}$, the difference between the deformation-constrained energy of the fragment and its equilibrium energy at the same temperature. This relation then defines $E^{*,int}$, and the energy dissipated along the the particular fission trajectory is determined by
\begin{equation}
E^{*,dis} = E^{*,int} - E^{*}_{FS}\; ,
\label{eq:en_TXE}
\end{equation}
where $E^{*}_{FS}$ is the initial excitation energy of the fissioning system.

For the asymmetric fission trajectories 1 and 2, starting at $T_{init}=0.8$ MeV, the dissipation energies $E^{*,dis}$ are $9.56$ and $11.00$ MeV, respectively, and the corresponding temperatures at scission are $T_{sci}=0.82$ and $T_{sci}=0.90$ MeV. The sum of the deformation energies of the resulting fission fragments $E^{*,def}=E^{*,def}_1+E^{*,def}_2$ is $7.98$ MeV for trajectory 1, and $16.85$ MeV for trajectory 2. The prescission kinetic energies $E^{k,pre}$ are $12.19$ and $12.56$ MeV for trajectory 1 and 2, respectively, reflecting the fact that only a portion of the potential energy difference between the initial point and the system at scission, is actually converted into collective flow energy.

\begin{figure}[!htbp]
\centering
\includegraphics[width=0.9\textwidth]{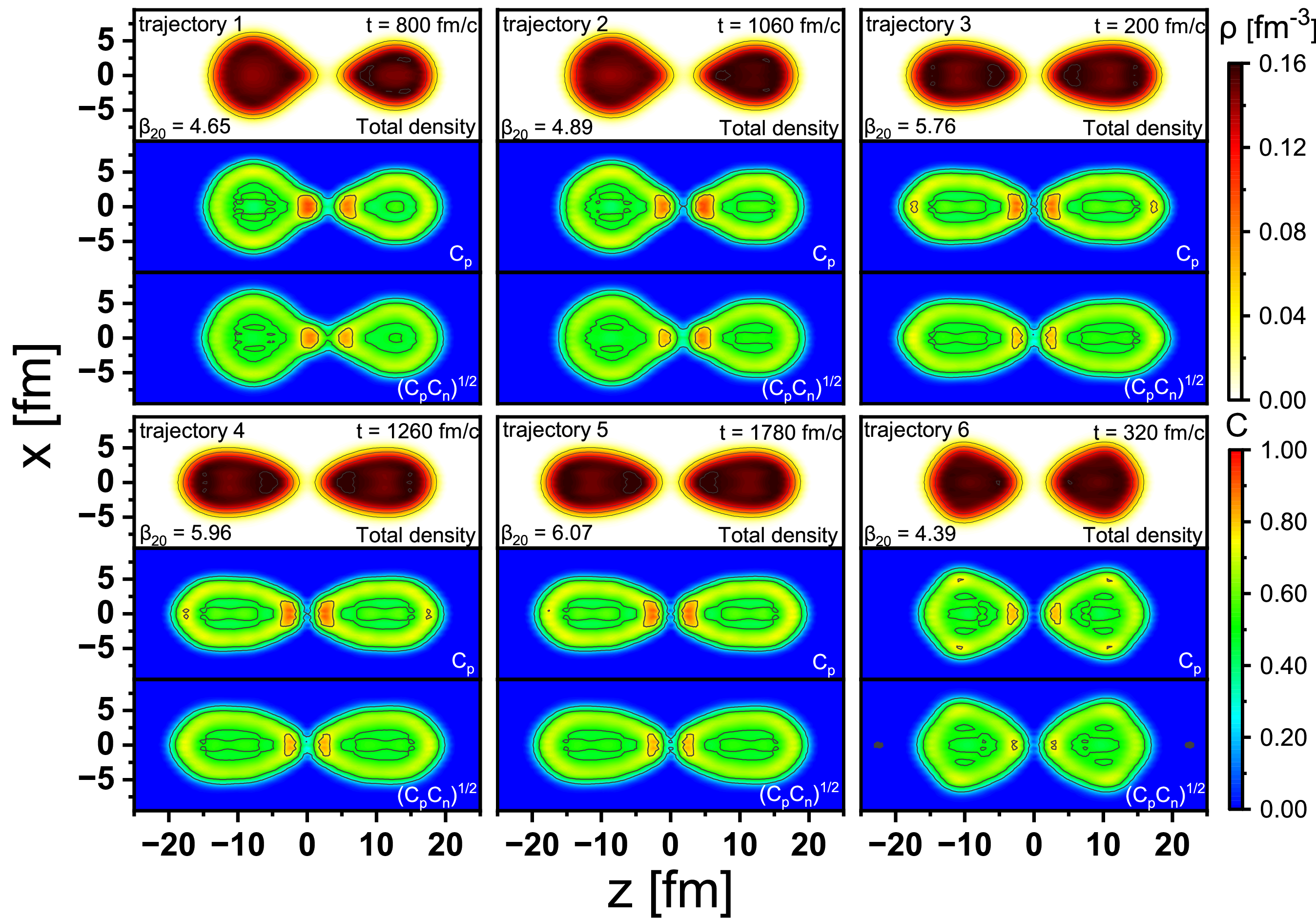}
\caption{Density profiles (color code in fm$^{-3}$) in the $x-z$ coordinate plane, and the corresponding proton $C_p$, and total $\sqrt{C_p C_n}$ localization functions, respectively, for trajectories 1 - 6, at times immediately after the scission event.}
 \label{fig:density_protile}
\end{figure}

We note that the definition of the specific instant at which scission occurs is somewhat arbitrary. One could, for instance, define it as a time at which the density in the neck region between the fragments falls below a chosen value. Another possibility is the moment when the two fragments separate just beyond the range of the nuclear attraction, so that the only interaction between them is the Coulomb repulsion. This choice is illustrated in the top panels of Fig.~\ref{fig:density_protile}, where we plot the density profiles of the fragments immediately after the scission event ($20$ fm/c after the density in the neck vanishes). Note, however, that the fragments are still entangled, as shown by the plots of the localization functions in the middle and bottom panels, to be defined below. For the second choice, we can compute the deformation energy of the fragments separately, and use Eq.~(\ref{eq:en_INT}) to calculate the total intrinsic excitation energy $E^{*,int}$ which, in this case, does not include a contribution of the interaction between the fragments.

\begin{figure}[!htbp]
\centering
\includegraphics[width=0.8\textwidth]{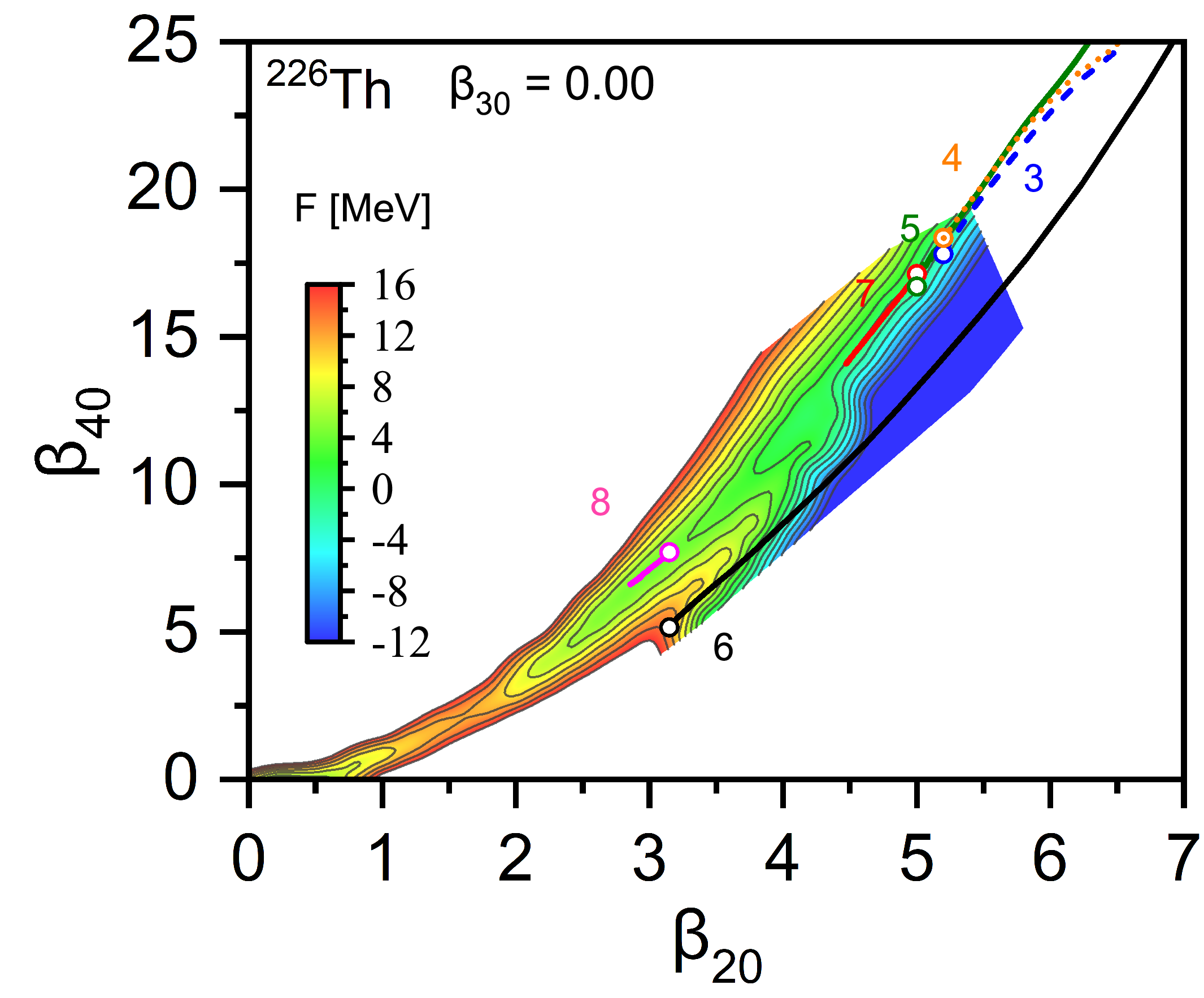}
\caption{Self-consistent deformation surface of Helmholtz free energy $F=E(T)-TS$ of $^{226}$Th in the $\beta_{20} - \beta_{40}$ plane of quadrupole-hexadecapole axially-symmetric deformation parameters,
calculated with the relativistic density functional PC-PK1 at the temperature $T=0.8$ MeV. Contours join points on the surface with the same energy (in MeV), and the contour interval is 2 MeV.
The curves denote the finite temperature TDDFT trajectories for six initial points on the energy surface.}
 \label{fig:ES_beta20_beta40_sym}
\end{figure}

These results are consistent with the ones we recently obtained for asymmetric fission trajectories of $^{234}$U, $^{240}$Pu, $^{244}$Cm, and $^{250}$Cf in Ref.~\cite{Li2023PRC_FTTDDFT}. In the case of $^{226}$Th, considered in the present analysis, the charge yields also display a pronounced contribution of symmetric fragment configurations \cite{Schmidt2000NPA,Chatillon2019PRC}. This is particularly interesting, because for symmetric fission the octupole moment for the whole system is zero at scission. For reflection-symmetric initial configurations, the octupole deformation $\beta_{30} = 0$  along fission trajectories that start after the barrier. Instead, in addition to the quadrupole elongation, we consider the hexadecapole $\beta_{40}$ deformation as the second collective coordinate. Close to scission, $\beta_{40}$ provides a measure of the thickness of the neck between two emerging fragments. For $\beta_{30} = 0$, in Fig.~\ref{fig:ES_beta20_beta40_sym} we plot the deformation energy surface of $^{226}$Th in the $\beta_{20} - \beta_{40}$ plane. Compared to the quadrupole-octupole energy surface (cf. Fig.~\ref{fig:ES_beta20_beta30}), the $T=0.8$ MeV quadrupole-hexadecapole Helmholtz free energy displays a more narrow valley. For $\beta_{20} \leq 3$, on both sides of the potential basin the energy increases very steeply, and the blue region of low energy for $\beta_{20} \geq 4$ corresponds to the fission valley. We have chosen six initial points for reflection-symmetric trajectories on this energy surface. When the trajectories start at $\beta_{30}=0$, the self-consistent time-evolution preserves reflection symmetry. Two trajectories, 7 for $\beta_{20}= 5.00$, and 8 for $\beta_{20} = 3.15$, start in the valley and do not lead to scission. Instead, they evolve the system back towards equilibrium, and eventually get trapped in local minima.

The trajectories 3 and 4 start at the same quadrupole deformation $\beta_{20}=5.20$, but at different hexadecapole coordinates: $\beta_{40}=17.80$ and $\beta_{40}=18.34$, respectively. Even though these two reflection-symmetric trajectories remain rather close during the time-evolution, the time it takes to scission along these fission trajectories differs dramatically, as shown in Fig.~\ref{fig:entropy_temperature}. While for trajectory 3 it takes only $\approx 200$ fm/c to reach the scission point, on trajectory 4 the nucleus fissions only after 1260 fm/c. Another interesting observation is that the temperature does not increase toward scission, as in the case of asymmetric fission, but in fact it can even exhibit a steep decrease just before scission, as illustrated for trajectory 4. The corresponding deformation, excitation, and dissipation energies at scission are listed in Table \ref{tab:dissipation}.

\begin{figure}[!htbp]
\centering
\includegraphics[width=0.6\textwidth]{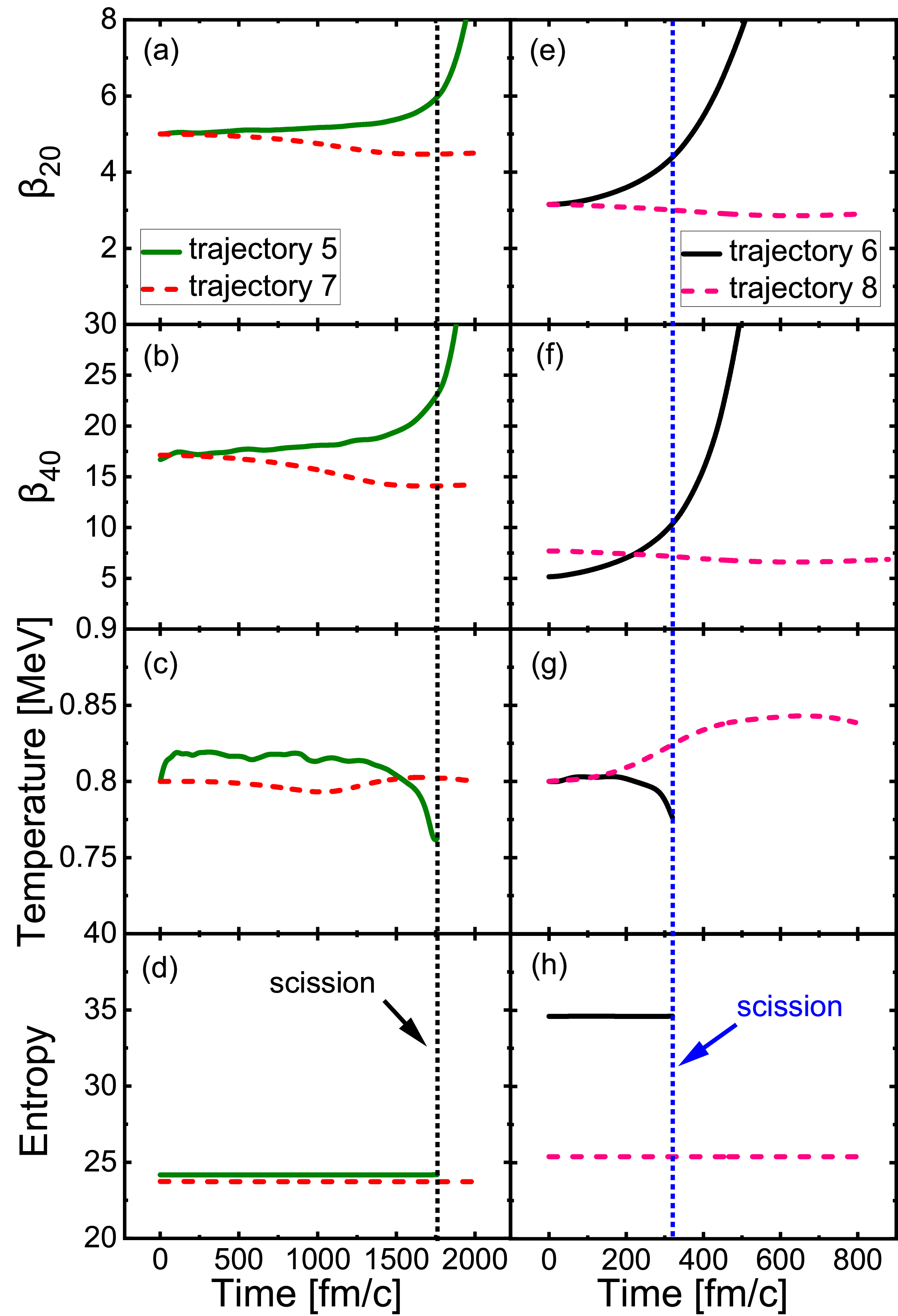}
\caption{Time-evolution of the quadrupole deformation $\beta_{20}$ (a), hexadecapole deformation $\beta_{40}$ (b), local temperature (c) and entropy (d) for trajectories 5 and 7 (left), and in panels (e) - (h) for trajectories 6 and 8 (right). }
 \label{fig:beta40_effect}
\end{figure}

The pairs of trajectories: 5 and 7, and 6 and 8, start at the same quadrupole deformation ($\beta_{20}= 5.00$ for 5 and 7, and $\beta_{20}= 3.15$ for 6 and 8) but different values of the hexadecapole parameter $\beta_{40}$. While
the initial $\beta_{40}$ of trajectory 5 is constrained to $16.70$, the initial state of trajectory 7 is located at a local minimum in $\beta_{40}$. When the constraints are released, the nucleus evolves to scission along the trajectory 5, while it gets trapped in a local minimum along the trajectory 7. The time evolution of $\beta_{20}$, $\beta_{40}$, the local temperature, and entropy are shown in panels (a) - (d) on the left of Fig.~\ref{fig:beta40_effect}, respectively. One notices that the
temperature for trajectory 5 slightly increases in the initial phase, but then decreases steeply near scission.

For the trajectories 6 and 8, the initial point $(\beta_{20},\beta_{30}) = (3.15,0.00)$ on the quadrupole-octupole energy surface shown in Fig.~\ref{fig:ES_beta20_beta30}, is very far from the scission contour. The two trajectories, however,
start from different points in the $\beta_{20} - \beta_{40}$ plane, as shown in Fig.~\ref{fig:ES_beta20_beta40_sym}. When the constraints are released for the time evolution, trajectory 8 gets confined at a local minimum in the valley, while trajectory 6 rapidly proceeds toward scission. The time evolution of the deformation parameters, temperature, and entropy are plotted in panels (e) - (h) on the right of Fig.~\ref{fig:beta40_effect}, respectively. Also in this case, it appears that temperature slightly decreases before scission and we note that, even though it starts at a much smaller quadrupole deformation, trajectory 6 leads to fission in a much shorter time compared to trajectory 5. As explained above, for the trajectories that end in fission, we can only follow the temperature and entropy up to scission because, after separating, the two fission fragments will generally have different temperatures.

For the symmetric trajectories 3, 4, and 5 with elongated configurations at scission (Fig.~\ref{fig:density_protile}), the deformation energies of fragments $E^{*,def}=E^{*,def}_1+E^{*,def}_2$ are $17.76$, $18.02$, and $18.02$ MeV, respectively. The prescission kinetic energies $E^{k,pre}$ are about $6$ MeV, and the dissipation energies along these trajectories are also rather small, as shown in Table \ref{tab:dissipation}. The resulting differences between the temperature at scission point $T_{sci}$ and the initial temperature $T_{init}$ are negative. This means that the fissioning system does not heat up on these trajectories to scission, the initial excitation energy is instead converted into deformation energy of the elongated configurations. For the symmetric trajectory 6 with compact shapes at scission, the initial energy is much higher than those for other trajectories (cf. Tab.~\ref{tab:dissipation}). The Coulomb energy at scission is larger because of the shorter distance between the two charge distributions. Since the energy surface (Fig.~\ref{fig:ES_beta20_beta40_sym}) around the initial point is very steep, the corresponding prescission kinetic energy is much larger than those corresponding to the other three symmetric fission trajectories with elongated configurations. The temperature at scission $T_{sci}$ is slightly smaller than the initial temperature $T_{init}$, even though the dissipation energy $E^{*,dis}=9.91$ MeV is very large compared to symmetric trajectories with elongated configuration at scission.

The density profiles immediately after scission, for trajectories 1 - 6, are shown in the corresponding top panels of Fig.~\ref{fig:density_protile}, respectively. For asymmetric fission (trajectories 1 and 2), the quadrupole deformations at scission are generally smaller than for elongated symmetric configurations (trajectories 3, 4, and 5), but comparable to those obtained for compact symmetric configurations (trajectory 6).
Well below saturation density, nuclear matter becomes inhomogeneous and a low-density system can locally minimize its energy by forming light clusters, in particular strongly bound $\alpha$-particles \cite{typel10,zinner13,ebran14,ebran17}.
The formation of clusters in the low-density neck region of a fissioning nucleus is intriguing  \cite{wuenschel14,ropke21,denisov21,Ren_22PRL,Li2023PRC_FTTDDFT},
and can be manifested by the kinematics of ternary fission events in which not only $^{4}$He, but also $^{3}$H and $^{6}$He cluster emission is observed.
In the TDDFT study of the final phase of the fission process that precedes scission \cite{Ren_22PRL},
we have shown that the mechanism of neck formation and its rupture are characterized by the dynamics of light clusters.
Subsequently, this mechanism has also been analyzed at finite temperature in Ref.~\cite{Li2023PRC_FTTDDFT}.
In a mean-field analysis, however, one cannot directly identify few-nucleon clusters and, as shown in Refs.~\cite{Ren_22PRL,Li2023PRC_FTTDDFT},
the one-body density at the time of scission does not exhibit signatures of cluster formation.
One must rather consider the corresponding time-dependent nucleon localization functions \cite{becke90,reinhard11}:
\begin{equation} \label{nlf}
C_{q\sigma}(\vec{r})=\left[1+\left(\frac{\tau_{q\sigma}\rho_{q\sigma}-{1\over 4}|\vec{\nabla}\rho_{q\sigma}|^2-\vec{j}^2_{q\sigma}}{\rho_{q\sigma}\tau^\mathrm{TF}_{q\sigma}}\right)^2\right]^{-1} ,
\end{equation}
for the spin $\sigma$ ($\uparrow$ or $\downarrow$)  and isospin $q$ ($n$ or $p$) quantum
numbers. $\rho_{q\sigma}$, $\tau_{q\sigma}$, $\vec{j}_{q\sigma}$, and $\vec{\nabla}\rho_{q\sigma}$ denote the nucleon density, kinetic energy density, current density, and density gradient, respectively.
$\tau^\mathrm{TF}_{q\sigma}={3\over 5}(6\pi^2)^{2/3}\rho_{q\sigma}^{5/3}$ is the Thomas-Fermi kinetic energy density.
The nucleon localization function is
derived by considering the conditional probability of finding a nucleon within a distance $\delta$ from a given nucleon
at point $\vec{r}$ with the same spin $\sigma$ ($=\uparrow$ or $\downarrow$)  and isospin $q$ ($=n$ or $p$) quantum
numbers. A small conditional probability indicates a high degree of localization of the reference nucleon.

For homogeneous nuclear matter $\tau = \tau^\mathrm{TF}_{q\sigma}$, the second and third term in the numerator vanish,
and $C_{q\sigma} = 1/2$. In the other limit $C_{q\sigma} (\vec{r}) \approx 1$
indicates that the probability of finding two nucleons with the same spin and isospin at the neighborhood of point $\vec{r}$ is very small.
This is the case for the $\alpha$-cluster of four particles: $p \uparrow$,  $p \downarrow$, $n \uparrow$,
and $n \downarrow$, for which all four nucleon localization functions $C_{q\sigma} \approx 1$.

For the asymmetric trajectories 1 and 2, the localization functions $C_p$ and $\sqrt{C_pC_n}$, shown in the middle and bottom panels of Fig.~\ref{fig:density_protile}, display values 0.4 -- 0.6 in the central regions of the fragments, which is consistent with homogeneous nuclear matter.
In the neck region between two fragments, however, the localization functions are close to 1, characteristic for the formation of $\alpha$ clusters. At scission the $\alpha$-like clusters are repelled by the Coulomb interaction, and are absorbed by the emerging fragments, where they induce strongly damped dipole oscillations along the fission axis.
In our previous studies, we only considered the mechanism of cluster formation in the low-density neck region for asymmetric fission trajectories.
In the case of induced fission of $^{226}$Th, for the symmetric fission trajectories 3, 4, and 5 with elongated scission configurations, we also find clear signatures of $\alpha$-like cluster formation in the neck region.
For the symmetric trajectory 6 which leads to a compact scission configuration with a shorter neck, the localization functions are less pronounced in the region between the fragments.

\subsection{Entropy of fragments and entanglement at finite temperature}
\begin{table}[!htbp]
  \caption{Entropies of the total system and fragments at scission, and entanglement between fragments, for the fission trajectories 1 - 6 of $^{226}$Th. }
  \label{Tab_ent}
  \bigskip
  \centering
  \renewcommand\arraystretch{1.5}
  \begin{tabular}{ccccccc}
    \hline\hline
      nucleus &\multicolumn{6}{c}{$^{226}$Th}\\
    \hline
      trajectory & 1 & 2 & 3 & 4 & 5 & 6 \\

     \hline

     $S_{tot}$                & 21.54    & 26.26    & 26.48    &  23.73   &   24.17  &  34.61       \\

     $S_1    $                & 14.20    & 16.55    & 14.19    &  12.98   &   13.33  &  18.68       \\

     $S_2    $                & 11.52    & 13.92    & 14.19    &  12.98   &   13.33  &  18.68       \\

     $L    $                  & 4.18     & 4.21     & 1.90     &  2.23    &   2.49   &  2.75       \\

     \hline
  \hline\hline
  \end{tabular}

\end{table}
As illustrated in Figs.~\ref{fig:entropy_temperature} and \ref{fig:beta40_effect}, the local microscopic entropy (\ref{Eq_entropy}) is constant along a trajectory leading to scission, as it should be for a unitary evolution of a closed system. However, at scission two fragments emerge and, in general, they will be characterized by different entropies. The level of their quantum entanglement is an interesting and open problem. The von Neumann entropy is generally defined in terms of the density matrix
\begin{equation}
S = - {\rm Tr} (\rho \ln \rho),
\end{equation}
or, in the case of separate fragments, in terms of the reduced density matrices $\rho_{\rm \sc 1} = {\rm Tr}_{\rm \sc 2} (\rho)$ and $\rho_{\rm \sc 2} = {\rm Tr}_{\rm \sc 1} (\rho)$,
and the traces are over the fragment $2$ and fragment $1$, respectively.
The von Neumann entropy of the fragments at $T=0$ MeV, can be obtained by the method introduced in Ref.~\cite{Klich2006JPA}.
For a compound nucleus at finite temperature $T$, however, the occupation probability of the single particle state $\psi_i$ is $f_i$ (cf. Eq.~(\ref{eq_fkt})), instead of $0$ or $1$ at $T=0$.
The von Neumann entropy $S^{(q)}_V$ for neutrons ($q=n$) or protons $(q=p)$ of the fragment, located in the subspace $V$, is evaluated by using the expression \cite{Klich2006JPA}
\begin{equation}
\begin{aligned}
   S_V^{(q)}&=-{\rm Tr} \{ {\bm M_V^{(q)}} \ln  {\bm M_V^{(q)}} + [{\bm I}- {\bm M_V^{(q)}}] \ln [{\bm I}- {\bm M_V^{(q)}}]\}\\
         &=-\sum_{i=1}^{N^{(q)}} \{ d_i^{(q)} \ln d_i^{(q)} + [1-d_i^{(q)}] \ln [1-d_i^{(q)}]\},
\end{aligned}
   \label{SV}
\end{equation}
where ${\bm I}$ is a unit matrix and $N^{(q)}$ is the number of neutron ($q=n$) or proton $(q=p)$ single particle levels of the total system.
$d_i$ are the eigenvalues of the overlap matrix $ {\bm M_V^{(q)}}$, whose elements are defined by the relation
\begin{equation}
   [{\bm M_V^{(q)}}]_{ij} =\sqrt{f_if_j} \langle \psi_i^{(q)} |\hat{\Theta}_V|\psi_j^{(q)}\rangle,
   \label{MV}
\end{equation}
where $\hat{\Theta}_V$ is the Heaviside function in coordinate space
\begin{equation}
\begin{split}
    \Theta_V(\bm r)=\left \{
\begin{array}{ll}
    1,     & {\rm if}~~\bm{r} \in V\\
    0,     & {\rm if}~~\bm{r} \not\in V
\end{array}
\right.
\end{split}.
\end{equation}
The mean-value of the nucleon number $N^{(q)}_V$, and its fluctuation $\Delta N^{(q)2}_V$ can also be obtained from the matrix $ {\bm M_V^{(q)}}$,
\begin{equation}
N^{(q)}_V={\rm Tr} [ {\bm M_V^{(q)}} ]=\sum_{i=1} d_i^{(q)},
\label{particle_numebr}
\end{equation}
\begin{equation}
\Delta N^{(q)2}_V={\rm Tr}\{ {\bm M_V^{(q)}}[{\bm I}- {\bm M_V^{(q)}}] \}=\sum_{i=1} d_i^{(q)}[1-d_i^{(q)}].
\label{particle_fluctuation}
\end{equation}
The entropy of the fragment in the subspace $V$ is the sum of its proton and neutron entropies,
\begin{equation}
S_V=S_V^{(n)}+S_V^{(p)}.
\end{equation}

Using the same method, one obtains the entropy $S_{\bar{V}}$ of the fragment in the complementary subspace $\bar{V}$.
The entanglement (mutual information) between the fragments can be defined \cite{Auletta2009Book}
\begin{equation}
L=S_V+S_{\bar{V}}-S_{tot},
\label{entanglement}
\end{equation}
where $S_{\rm tot}$ is the entropy of total system. Here all three entropies are determined at the scission point, that is at $T_{sci}$, and $S_{tot}$ is the entropy of the trajectory that leads to the formation of the particular pair of fragments.

The results are summarized in Table \ref{Tab_ent}, where we list, for the six fission trajectories of $^{226}$Th considered in the present study, the entropy of the compound system, the relative entropies of the resulting fragments, and the entanglement (\ref{entanglement}). For the asymmetric fission trajectories 1 and 2, the entropy of the heavy fragment is larger than the one of the light fragment, simply reflecting the size of the corresponding nucleus.
The entanglement between the two fragments is $\approx 4.2$, for both trajectories.
For symmetric fission (trajectories 3, 4, 5, and 6) the entropies of the two fragments are, of course, identical. The entanglement, however, appears to be significantly reduced in comparison to the two asymmetric fission events. It would be interesting to perform a more general study of the entanglement of fission fragments, as a function of fragment's charges and masses, as well as angular momenta (Shannon entropies). Such an analysis is beyond the scope of the present work, and will be the subject of a future study.

Finally, we calculate the dispersions of nucleon number to validate the results. The dispersions $\sqrt{\Delta N^{2}}$ for the total system can be obtained using Eq.~(\ref{particle_fluctuation}). The dispersions of the neutron number at scission are 1.32, 1.49, 1.51, 1.47, 1.49, and 1.64 for trajectories 1 - 6, respectively, which is only about 1.0\% of the total neutron number. The dispersions of the proton number at scission are 1.34, 1.29, 1.29, 1.24, 1.21, and 1.39 for trajectories 1 - 6, respectively, which is only about 1.5\% of the total proton number. Therefore, we are confident that the results are valid.

\section{Summary}\label{sec_summ}

The saddle-to-scission phase of the induced-fission process is characterized by a partial conversion of the potential energy difference between saddle and scission into collective flow, energy dissipation, and the formation of deformed, excited, and entangled fission fragments. To model this complex process, we have recently developed a microscopic finite-temperature model based on time-dependent nuclear density functional theory (TDDFT), that allows to follow the evolution of local temperature along fission trajectories.

In our previous study \cite{Li2023PRC_FTTDDFT}, characteristic asymmetric fission trajectories were considered in $^{240}$Pu, $^{234}$U, $^{244}$Cm, and $^{250}$Cf, and the partition of the total energy into various kinetic and potential energy contributions at scission analyzed. The focus of this work has been the difference between symmetric and asymmetric fission trajectories in  $^{226}$Th. For an initial temperature $T=0.8$ MeV,
which corresponds to the experimental peak energy for photo-induced fission \cite{Schmidt2000NPA}, the deformation energy surface of Helmholtz free energy $F=E(T)-TS$ is obtained with constrained microscopic self-consistent density mean-field calculations, using the relativistic density functional PC-PK1. The axially symmetric quadrupole, octupole, and hexadecapole deformations represent coordinates in the collective space. Starting from initial points that are located beyond the saddle, the dynamic model propagates nucleons along self-consistent isentropic trajectories toward scission. The non-equilibrium generalization of the chemical potential and temperature are adjusted at each step of the time-evolution, so that the particle number and total energy, respectively, are conserved along a TDDFT trajectory. Dissipation effects that characterize the saddle-to-scission dynamics, and the formation of excited and deformed fission fragments can be investigated.

As shown in our study of asymmetric fission trajectories of Ref.~\cite{Li2023PRC_FTTDDFT}, the increase in temperature between the initial point and scission is generally of the order of 10\% to 20\%. Only a portion of the potential energy difference between the initial and scission points is converted into collective flow, and the dissipated energy is of the order of $10$ MeV. An interesting and, at first glance, counter-intuitive result of the present study is that, for symmetric fission trajectories, near scission the local temperature does not necessarily increase and can even exhibit a steep decrease. In these cases dissipation does not lead to a temperature increase, but rather to a formation of strongly deformed symmetric fragment configurations. For elongated symmetric fragments the prescission kinetic energy is relatively small ($\approx 6$ MeV), and the energy dissipated along the corresponding trajectory is smaller ($\leq 5.5$ MeV), compared to asymmetric fission trajectories.  A compact symmetric scission configuration exhibits smaller deformation energy than elongated shapes, but much larger collective flow and dissipation energies.

A very interesting topic for future fission studies, and for which the present work has only outlined the framework, is the level of entanglement (mutual information) of fragments at scission. A preliminary result, obtained by analyzing the entropies of the compound fissioning system and the relative entropies of the fragments at scission, is that the entanglement for symmetric configurations appears to be significantly reduced in comparison to asymmetric fission events. In addition to the entanglement resulting from the von Neumann entropies computed with density operators, it will be interesting to analyze the Shannon entropy and entanglement of various observables that characterize fission fragments (charge, mass, angular momentum, excitation energy).

\begin{acknowledgments}
This work has been supported in part by the High-End Foreign Experts Plan of China,
the National Natural Science Foundation of China (Grants No.11935003 and No.12141501),
the State Key Laboratory of Nuclear Physics and Technology, Peking University (Grant No. NPT2023ZX03)
the High-Performance Computing Platform of Peking University,
by the project ``Implementation of cutting-edge research and its application as part of the Scientific Center of Excellence for Quantum and Complex Systems, and Representations of Lie Algebras'', PK.1.1.02, European Union, European Regional Development Fund, and by the Croatian Science Foundation under the project Relativistic Nuclear Many-Body Theory in the Multimessenger Observation Era (IP-2022-10-7773).

\end{acknowledgments}

\bigskip


%

\end{document}